\documentclass[onecolumn,aps,floatfix,pra,superscriptaddress,nofootinbib]{revtex4}
\usepackage{hyperref}
\usepackage{color}
\usepackage{graphicx}
\usepackage{bm}
\usepackage{amsmath}
\usepackage{cleveref}
\usepackage{enumerate}
\usepackage{upgreek}
\usepackage[subnum]{cases}

\newcommand{\cl}{ \text{cl} }

\newcommand{\pa}{ \partial }
\newcommand{\hb}{ \hbar }
\newcommand{\si}{ \sigma }

\newcommand{\ga}{ \gamma }

\newcommand{\del}{ \delta }
\newcommand{\al}{ \alpha }

\newcommand{\erfc}{ \text{erfc} }

\begin{document}
\title{Dissipative quantum backflow}
\author{S. V. Mousavi}
\email{vmousavi@qom.ac.ir}
\affiliation{Department of Physics, University of Qom, Ghadir Blvd., Qom 371614-6611, Iran}
\author{S. Miret-Art\'es}
\email{s.miret@iff.csic.es}
\affiliation{Instituto de F\'isica Fundamental, Consejo Superior de Investigaciones Cient\'ificas, Serrano 123, 28006 Madrid, Spain}
\begin{abstract}

Dissipative backflow is studied in the context of open quantum systems. This  theoretical analysis is carried out within two frameworks, the effective time-dependent Hamiltonian due to Caldirola-Kanai (CK) and the Caldeira-Leggett (CL) one where a master equation is used to describe the reduced density matrix in presence of dissipation and temperature of the environment. Two examples are considered, the free evolution of one and two Gaussian wave packets as well as  the time evolution under a constant field. Backflow is shown to be reduced with dissipation and temperature but never suppressed. Interestingly enough, quantum backflow is observed when considering both one and two Gaussian wave packets within the CL context. Surprisingly, in both cases, the backflow effect seems to be persistent at long times. Furthermore,  the constant force $ m g\geq 0 $ behaves against backflow. However, the classical limit of this quantum effect within the context of the classical Schr\"odinger equation is shown to be present. Backflow is also analyzed as an eigenvalue problem in the Caldirola-Kanai framework. In the free propagation case, eigenvalues are independent on mass, Planck constant, friction and its duration  but, in the constant force case, eigenvalues depend on a factor which itself is a combination of all of them as well as the force constant.

\end{abstract}

\maketitle



\section{Introduction}

The so-called quantum backflow is a very interesting nonclassical effect which is quite counter-intuitive. It happens when a free particle described by a one-dimensional wave function located in the negative axis of the coordinate, and 
consisting entirely of positive momenta, displays a non-decreasing probability of remaining in the negative region during certain periods of time. 
Since the first recognition by Allcock \cite{allcock} in 1969 when studying arrival times in quantum mechanics,  not too much attention has been paid 
in the literature. The first systematic study of this effect is due to Bracken and Melloy \cite{BrMe-JPA-1994} emphasizing that quantum 
backflow merely reflects the structure of the Schr\"odinger equation. These authors showed that the highest probability which can flow back 
from positive to negatives values of the coordinate is around 0.04 for a superposition of two wave planes with positive momenta.
Then they showed that the maximum amount of probability backflow that can occur in general over any finite time interval is about that value
 to be independent on the time interval, mass of the particle and Planck constant. 
Thus, a new dimensionless quantum number was then reported. Bracken and Melloy \cite{BrMe-AP-1998} then studied the effect in the presence 
of a constant field and also  relativistic particles obeying Dirac equation. They also showed that the probability flow can be regarded as an
eigenvalue problem of the flux operator. Optimization numerical problems were reported by Penz et  al. \cite{Penz-2006}

Quantum backflow is also connected to interference and quantum arrival times \cite{Mu-Lea-PR-2000}. This effect has also been related 
to superoscillations \cite{Berry-2006} and weak values \cite{Berry-2010}. More recently, Yearsley et al. \cite{Yearsley-2012, Yearsley-2013} 
have analyzed the classical limit and discussed some specific measurement models. Backflow can not take place for a single 
Gaussian wave packet but  it occurs for states consisting of superpositions of Gaussian wavepackets \cite{Yearsley-2012}. Albarelli et al. 
have addressed the notion of nonclassicality arising from the backflow effect and analyzed its relationship with the corresponding one on 
the negativity of the Wigner function \cite{Albarelli-2016}. 
Very recently, it was also argued that the backflow under the presence of a constant field is mathematically equivalent to the problem of 
diffraction in time for particles initially confined to a semi-infinite line, expanding in free space \cite{Gu-PRA-2019}.
As far as we know, no experimental evidence of this effect has been reported yet.

On the other hand, apart from a work studying the arrival time problem in the framework of decoherent
histories for a particle coupled to an environment \cite{Ye-PRA-2010}, we are not aware of any other study carried out in the context of open quantum 
systems; in particular, how the friction and temperature can influence the backflow effect. In this work, we address this issue within two 
different frameworks, the Caldirola-Kanai \cite{Caldirola-Kanai} and the Caldeira-Leggett \cite{CaLe-PA-1983, Caldeira-book-2014} approaches.
Backflow is shown to be reduced with dissipation and temperature but never suppressed. Interestingly enough, quantum backflow is 
surprisingly observed for a single Gaussian wave packet within the CL context. Furthermore,  the constant force $ m g\geq 0 $ behaves 
against backflow. However, the classical limit of backflow within the context of the classical Schr\"odinger equation is shown to be present. 
 
This work is organized as follows.  In Section II, an effective time dependent Hamiltonian, the so-called Caldirola-Kanai Hamiltonian, is used 
to describe dissipation. This approach has been considered several times as an effective way to tackle such dissipative problems. It has been 
widely shown that this approach provides acceptable results \cite{MoMi-AO-2018, Sanz-2014}. Free evolution of one and two 
Gaussian wavepackets and under the presence of a constant field are studied and analyzed. In Section III, the backflow is considered as 
an eigenvalue problem with existing dissipation in the CK framework for the free evolution and under the presence of a constant field. The classical 
limit is afterward discussed within the so-called classical Schr\"odinger equation \cite{Richardson-2014}.
In Section IV, the backflow effect is analyzed within the CL approach where a master equation is used to describe the reduced 
density matrix in terms of friction and temperature of the environment. Free evolution of one and two Gaussian wave packets as well as the dynamics 
under the presence of a constant field is again studied. In Section V, some numerical results are presented and discussed. Finally, in the last 
section, a summary and some conclusions are presented.

\section{The Caldirola-Kanai framework}

Among different approaches for dealing with dissipation in physical systems, one can mention effective time dependent Hamiltonians. 
Dissipation is taken into account via explicitly time-dependent Hamiltonians, thus avoiding to deal with the environment degrees of freedom. 
Canonical formalism is preserved in such an approach which can be used for constructing the quantum analogue of the corresponding 
dissipative dynamics. 
In this regard, the so-called CK model \cite{Caldirola-Kanai} can be mentioned as a paradigm in which a Hamiltonian 
formulation of the Langevin equation with zero fluctuations is used. This approach has been employed to study dissipative Bohmian trajectories
\cite{Sanz-2014,MoMi-JPC-2018} and dissipative tunneling \cite{MoMi-AO-2018}

For the sake of simplicity, in one dimension, the classical Langevin equation without fluctuating force reads as
\begin{eqnarray} \label{eq: Langevin}
m \ddot{x} + \eta \dot{x} + \frac{\pa V}{\pa x} &=& 0
\end{eqnarray}
where $ m $ is the mass of the particle, $ \eta $ is the damping constant and $ V(x) $ is the external potential. 
By defining the relaxation constant $\ga$ as \cite{CaLe-PA-1983}
\begin{eqnarray} \label{eq: relax_const}
\ga &=& \frac{\eta}{2m}
\end{eqnarray}
%
the Langevin equation (\ref{eq: Langevin}) can be derived from the time-dependent Lagrangian
\begin{eqnarray} \label{eq: CK_Lag}
\mathcal{L}_{\text{CK}} &=& \left( \frac{1}{2} m\dot{x}^2 -  V(x) \right) e^{2\ga t}
\end{eqnarray}
from which the  CK Hamiltonian 
\begin{eqnarray} \label{eq: CK_Lag}
\mathcal{H}_{\text{CK}} &=& \frac{p_c^2}{2m} e^{-2\ga t}  +  V(x) e^{2\ga t} 
\end{eqnarray}
is obtained, $p_c$ being the canonical momentum $p_c = \frac{\pa \mathcal{L}_{\text{CK}}}{\pa \dot{x}} = m \dot{q} e^{2\ga t}$. 
In quantum mechanics, one has to replace the canonical momentum $p_c$ by $-i\hb \pa_x$. In this way, the time-dependent Schr\"{o}dinger equation 
reads as
\begin{eqnarray} \label{eq: CK}
i \hb \frac{\pa }{\pa t} \psi(x, t)  &=& \bigg[ - e^{-2\ga t} \frac{\hb^2}{2m} \frac{\pa^2}{\pa x^2}  + e^{2\ga t} V(x) \bigg] \psi(x, t)  .
\end{eqnarray}
The probability current density $j(x, t)$ fulfilling the continuity equation
\begin{eqnarray} \label{eq: con_CK}
\frac{\pa |\psi(x, t)|^2}{\pa t} + \frac{\pa j(x, t)}{\pa x}  &=& 0 ,
\end{eqnarray}
is given by
\begin{eqnarray} \label{eq: pcd_CK}
j(x, t) &=& \frac{\hb}{m} \text{Im} \left\{ \psi^* \frac{\pa \psi}{\pa x}  \right\} e^{-2\ga t}   .
\end{eqnarray}

In the following we first show that backflow does not take place for a single Gaussian wave packet. Then we study the effect of superposition 
of two Gaussian wave packets both in free propagation and in the presence of a constant field.  

\subsection{Free propagation of a Gaussian wave packet}

For the initial momentum-space wave function
\begin{eqnarray} \label{eq: fourier}
\phi(p, 0) &=& \frac{1}{(2\pi \si_p^2)^{1/4}} \exp \left[ - \frac{(p-p_0)^2}{4\si_p^2} \right]
\end{eqnarray}
the configuration-space wavefunction reads
\begin{eqnarray} \label{eq: Gauss0}
\psi(x, 0) &=& \frac{1}{( 2\pi )^{1/4}} \sqrt{\frac{2\si_p}{\hb}} \exp \left[ - \frac{\si_p^2 ~ x^2}{\hb^2} + i \frac{p_0}{\hb} x \right] .
\end{eqnarray}
The solution of the CK equation (\ref{eq: CK}) for free propagation of the Gaussian wave function (\ref{eq: Gauss0}) is given by \cite{MoMi-JPC-2018}
\begin{eqnarray} \label{eq: Gausst}
\psi(x, t) &=& \frac{1}{(2\pi s_t^2)^{1/4}} \exp \left[ - \frac{\si_p (x-x_t)^2}{2\hb s_t} + i \frac{p_0}{\hb} (x-x_t) + \frac{i}{\hb} \mathcal{A}_{\cl}(t) \right]
\end{eqnarray}
where, $s_t$, $x_t$ and $\mathcal{A}_{\cl}(t) \equiv \mathcal{A}_{\cl,t}$ are the complex width, the center of the wavepacket which follows a classical trajectory and 
the classical action expressed as
\begin{numcases}~
s_t = \frac{1}{2\si_p} \left( \hb + i \frac{ 2\si_p^2}{m} \frac{1-e^{-2\gamma t}}{2\gamma} \right) ,  \\
x_t = \frac{p_0}{m} \frac{1-e^{-2\gamma t}}{2\gamma} , \label{eq: xt} \\
\mathcal{A}_{\text{cl},t} = \frac{p_0^2}{2m} \frac{1-e^{-2\gamma t}}{2\gamma} ,
\end{numcases}
respectively.

With respect to Eq. (\ref{eq: fourier}), the probability of obtaining a negative value in a measurement of the momentum is given by
\begin{eqnarray} \label{eq: negative-prob}
\text{Prob}(p<0) &=& \int_{-\infty}^0 dp | \phi(p, 0) |^2 = 
\frac{1}{2} \erfc \left[ \frac{p_0}{\sqrt{2} \si_p} \right] .
\end{eqnarray}
By a proper choice of the initial parameters $p_0$ and $\si_p$, one can easily make the value of $ \text{Prob}(p<0) $  negligibly small. In this way,
one makes sure that the wave packet (\ref{eq: Gauss0}) has only positive momentum components. From Eq.  (\ref{eq: Gausst}), 
the probability that the particle remains in the half-space $ x<0 $ is
\begin{eqnarray} \label{eq: prob_x<0}
P(t) &=& \int_{-\infty}^0 dx |\psi(x, t)|^2 = \frac{1}{2} \erfc \left[ \frac{x_t}{\sqrt{2} \si_t} \right]
\end{eqnarray}
where 
\begin{eqnarray} \label{eq: sigmat}
\si_t &=& |s_t| = \frac{1}{2\si_p} \sqrt{ \hbar^2 + \frac{ 4 \si_p^4 }{ m^2 } \left( \frac{1-e^{-2\gamma t}}{2\gamma} \right)^2 }
\end{eqnarray}
being the width of the probability density  $ |\psi(x, t)|^2 $.
%
Noting that the complementary error function is a decreasing function of its argument and that the argument is an increasing function of time i.e., $ \frac{d}{dt} ( \frac{x_t}{\sqrt{2} \si_t} ) > 0 $ for $ p_0 > 0 $. Thus, $P(t)$ is therefore a decreasing function of time and the backflow 
effect is not expected to occur for a single Gaussian wave packet.

\subsection{Dissipative backflow for the superposition of two Gaussian wave packets}

Now consider the momentum representation of the initial state as  
\begin{eqnarray} \label{eq: wf0_momentum}
\phi(p, 0) &=& N \frac{1}{(2\pi \si_p^2)^{1/4}} \left\{
\exp \left[ - \frac{(p-p_{0a})^2}{4\si_p^2} \right]
+ \al e^{i\theta} \exp \left[ - \frac{(p-p_{0b})^2}{4\si_p^2} \right]
\right \}, 
\end{eqnarray}
where $N$, the normalization constant, $\al$ and $\theta$ are all real numbers and
\begin{eqnarray} \label{eq: N_cons}
N &=& \left( 1 + \al^2 + 2 \al e^{- (p_{0a}-p_{0b})^2 / 8 \si_p^2} \cos \theta \right)^{-1/2}    .
\end{eqnarray}
State (\ref{eq: wf0_momentum}) is a superposition of two Gaussian wave packets in momentum space with the same width $\si_p$ but different 
centers $p_{0a}$ and $p_{0b}$. When a quantum system is described by Eq. (\ref{eq: wf0_momentum}) in momentum space, the probability 
for obtaining a negative value in a measurement of momentum is
\begin{eqnarray} \label{eq: prob_p<0}
\text{Prob}(p<0) &=& \int_{-\infty}^0 | \phi(p, 0) |^2 
\nonumber \\
&=& 
\frac{1}{2} N^2 \bigg\{ 
\erfc \left[  \frac{p_{0a} }{ \sqrt{2} \si_p } \right] 
+ \alpha^2 \erfc \left[ \frac{p_{0b} }{ \sqrt{2} \si_p } \right]
+ 2 \al ~e^{- (p_{0a}-p_{0b})^2 / 8\si_p^2} \cos \theta  ~ \erfc \left[ \frac{(p_{0a}+p_{0b})}{ 2\sqrt{2}\si_p } \right] \bigg\}     .
\end{eqnarray}
The initial wave function in configuration space is obtained by taking the Fourier-inverse transform of Eq. (\ref{eq: wf0_momentum}) and reads
\begin{eqnarray} \label{eq: wf_sup_0_Gauss}
\psi(x, 0) &=& N \frac{1}{( 2\pi )^{1/4}} \sqrt{\frac{2\si_p}{\hb}} (e^{i p_{0a} x/\hb} + \al e^{i\theta} e^{i p_{0b} x/\hb} )~ e^{-\si_p^2 x^2/\hb^2} 
\equiv N( \psi_a(x, 0) + \al e^{i\theta} \psi_b(x, 0) )   .
\end{eqnarray}
Due to the linearity of the CK equation, the time evolution of this state is given by 
\begin{eqnarray} \label{eq: wf_sup_t}
\psi(x, t) &=& N( \psi_a(x, t) + \al e^{i\theta} \psi_b(x, t) )   .
\end{eqnarray}
Thus, the {\it free} evolution of the probability density of the state given by Eq. (\ref{eq: wf_sup_0_Gauss}) can be expressed as
\begin{eqnarray} \label{eq: psi2_t}
|\psi(x, t)|^2 & = & N^2 \frac{1}{\sqrt{2\pi} \si_t}
\bigg\{
\exp \left[ - \frac{ ( x - x_{ta})^2  }{ 2 \si_t^2 } \right] 
+ \alpha \exp \left[ d_1 - \frac{ ( x - d_2(t))^2  }{ 2 \si_t^2 } \right] 
+ \alpha \exp \left[ d^*_1 - \frac{ ( x - d^*_2(t))^2  }{ 2 \si_t^2 } \right] 
\nonumber \\
&~& \qquad \qquad
+~ \alpha^2 \exp \left[ - \frac{ ( x - x_{tb})^2  }{ 2 \si_t^2 } \right] 
\bigg \}
\end{eqnarray}
where $x_{ta}$ and $x_{tb}$ are the centers of wave packets and
\begin{numcases}~
d_1 = - i \theta - \frac{(p_{0a} - p_{0b})^2 }{8\si_p^2} \\
d_2(t) = \frac{ x_{ta} + x_{tb} }{2}  + i \frac{\hb(p_{0a} - p_{0b}) }{4\si_p^2} 
\end{numcases}
with $p_{0a}$ and $p_{0b}$ being the initial momenta and $\si_t$ is given by Eq. (\ref{eq: sigmat}). 
From the probability density (\ref{eq: psi2_t}), one can easily write 
\begin{eqnarray} \label{eq: prob_x<0_CK}
P(t) &=& \frac{1}{2} N^2 \bigg\{ 
\erfc \left[ \frac{ x_{ta} }{\sqrt{2} \si_t}  \right]
+ \al^2 \erfc \left[ \frac{ x_{tb} }{\sqrt{2} \si_t} \right]
\nonumber \\
&+& 
2 \al e^{-(p_{0a} - p_{0b})^2 / 8\si_p^2 } 
\bigg( \cos\theta~ \text{Re} \left\{ \erfc \left[ \frac{d_2(t)}{\sqrt{2}\si_t} \right]
\right\} 
+ \sin\theta~ \text{Im} \left\{ \erfc \left[ \frac{d_2(t)}{\sqrt{2} \si_t} \right]
\right\} \bigg)
\bigg \}
\end{eqnarray}
for the probability of remaining in the region $x<0$. 
By noting the positive values for kick momenta, one observes that the arguments of the first two complementary error functions of the above 
equation are increasing functions of time. Thus, the last two terms are responsible for quantum backflow. 


\subsection{Dissipative backflow for the superposition of two Gaussian wave packets under the presence of a constant field}

On the other hand, the evolution of the pure state (\ref{eq: wf_sup_0_Gauss}) under Eq. (\ref{eq: CK}) and in the presence of the linear potential
\begin{eqnarray} \label{eq: lin_pot}
V(x) &=& - m g x
\end{eqnarray}
yields the same Eq. (\ref{eq: prob_x<0_CK}) for the probability of remaining in the region $x<0$ but with the replacement of $x_t$ by 
$ q_t $ where
\begin{eqnarray} \label{eq: xt_g} 
q_t &=& \frac{p_0}{m} \frac{1-e^{-2\gamma t}}{2\gamma} + g \frac{2\ga t-1+e^{-2\gamma t}}{4\gamma^2} 
\end{eqnarray}
is the classical trajectory in the presence of the constant force $ m g $.

\section{Probability flow as an eigenvalue problem in the CK framework and the classical limit}

\subsection{ Free evolution}

Bracken and Melloy \cite{BrMe-JPA-1994} constructed an eigenvalue equation for the probability flow. Here, following the same steps as the 
previous authors, we are going to build a similar eigenvalue equation but in the dissipative CK context. From the continuity equation 
(\ref{eq: con_CK}), one writes that
\begin{eqnarray}
\frac{d}{dt} \int_{-\infty}^0 dx |\psi(x, t)|^2 &=& - \int_{-\infty}^0 dx \frac{\pa j(x, t)}{\pa x} = -j(0, t)
\end{eqnarray}
where in the second equality we have used the square-integrability of the wave function. Thus, for the backflow in the time interval $[0, \tau]$ 
one obtains that 
\begin{eqnarray} \label{eq: P_j}
\Delta_P &\equiv& P(\tau) - P(0) = - \int_0^{\tau} dt ~ j(0, t)
\end{eqnarray}
the wave function at the instant $t$ being related via the following integral equation to the initial wave function
\begin{eqnarray} \label{eq: wft_wv0}
\psi(x, t) &=& \int_{-\infty}^{\infty} dx' G(x, t; x', 0) \psi(x', 0)
= \int_{-\infty}^{\infty} dx' G(x, t; x', 0) \frac{1}{ \sqrt{2\pi \hb} } \int_0^{\infty} dp' e^{i p' x' / \hb} \phi(p')
\end{eqnarray}
where in the second equality the Fourier transform of the initial wave function involves only positive momenta and $G(x, t; x', 0)$ is the propagator 
for free evolution \cite{MoMi-JPC-2018},
\begin{eqnarray} \label{eq: propagator}
G(x, t; x', 0) &=& \sqrt{ \frac{m}{2\pi i \hb ~ \uptau(t)} } \exp \left[ \frac{im (x-x')^2}{2\hb ~ \uptau(t)}  \right] 
\end{eqnarray}
with
\begin{eqnarray} \label{eq: uptau}
\uptau(t) &=& \frac{1-e^{-2\ga t}}{2\ga}    .
\end{eqnarray}
From Eqs. (\ref{eq: P_j}), (\ref{eq: wft_wv0}) and (\ref{eq: propagator}), we have that
\begin{eqnarray} \label{eq: DelP}
\Delta_P &=& \int_0^{\infty} \int_0^{\infty} dp dq ~ \phi^*(p) K(p, q) \phi(q)
\end{eqnarray}
where
\begin{eqnarray} \label{eq: K}
K(p, q) &=& - \frac{ p + q }{4\pi m \hb} \int_0^{\tau} dt ~  \exp \left[ i \frac{p^2-q^2}{2m\hb} \uptau(t) \right]   \nonumber \\
&=& -\frac{1}{\pi} \frac{1}{p-q} \exp \left[ i \uptau(\tau) \frac{p^2-q^2}{4m\hb} \right]
\sin \left( \uptau(\tau)\frac{ p^2 - q^2 }{4m\hb} \right)     .
\end{eqnarray}
In order to optimize $\Delta_P$ in Eq. (\ref{eq: DelP}) under the constraint 
\begin{eqnarray} \label{eq: constraint}
\int_0^{\infty} dp |\phi(p)|^2 &=&  1
\end{eqnarray}
the method of Lagrange multipliers is used to construct the functional
\begin{eqnarray} \label{eq: functinal}
I[\phi, \phi^*] &=&  \int_0^{\infty} \int_0^{\infty} dp dq ~ \phi^*(p) K(p, q) \phi(q) - \lambda \int_0^{\infty} dp ~ \phi(p)^* \phi(p)
\end{eqnarray}
where $ \lambda $ is the Lagrange multiplier. From the optimization condition 
$ \del I = I[\phi, \phi^* + \del \phi^*] - I[\phi, \phi^*] = 0  $, 
the eigenvalue equation is then expressed as 
\begin{eqnarray} \label{eq: eigenvalue}
\int_0^{\infty} dq ~ K(p, q) \phi(q) &=&  \lambda \phi(p)
\end{eqnarray}
which in its standard form should be written as 
\begin{eqnarray} \label{eq: eigeneq}
\int_0^{\infty} dq ~ K(p, q) \phi_{\lambda}(q) &=&  \lambda \phi_{\lambda}(p)   .
\end{eqnarray}
From the property
\begin{eqnarray} \label{eq: K_property}
K^*(p, q) &=& K(q, p)
\end{eqnarray}
we prove, in the following, that eigenvalues are real and eigenfunctions corresponding to distinct eigenvalues are orthogonal i.e., the integral operator $K$ is Hermitian.  
By multiplying the complex-conjugated eigenfunction $\phi^*_{\lambda'}(p)$ in the eigenvalue equation 
corresponding to the eigenvalue $\lambda'$, multiplying $\phi_{\lambda}(p)$ in the complex-conjugated eigenvalue equation
corresponding to the eigenvalue $\lambda^*$, subtracting the resulting equations and finally integrating over $p$, one has that
\begin{eqnarray}
\int_0^{\infty} dp \int_0^{\infty} dq  
\bigg\{ \phi^*_{\lambda'}(p) K(p, q) \phi_{\lambda}(q) - \phi_{\lambda}(p) K^*(p, q) \phi^*_{\lambda'}(q)  \bigg\} 
&=& (\lambda - \lambda'^*) \int_0^{\infty} dp ~ \phi^*_{\lambda'}(p) \phi_{\lambda}(p)   .
\end{eqnarray}
Now, by using Eq. (\ref{eq: K_property}) and interchanging $q$ by $p$ and viceversa in the second term of the left-hand-side, 
this term will be the same as the first one. Thus, we have 
\begin{eqnarray}
(\lambda - \lambda'^*) \int_0^{\infty} dp ~ \phi^*_{\lambda'}(p) \phi_{\lambda}(p) &=& 0   .
\end{eqnarray}
If we set $ \lambda' = \lambda $, $ \lambda - \lambda^* = 0 $ revealing that the eigenvalues are real. Then,
\begin{eqnarray}
(\lambda - \lambda') \int_0^{\infty} dp ~ \phi^*_{\lambda'}(p) \phi_{\lambda}(p) &=& 0 
\end{eqnarray}
which for $ \lambda' \neq \lambda $ yields the orthogonality of eigenfunctions corresponding to different eigenvalues.

From Eqs. (\ref{eq: eigenvalue}) and (\ref{eq: DelP}) and noting the constraint (\ref{eq: constraint}) one has 
\begin{eqnarray} \label{eq: DelP_lam}
\Delta_P &=&  \lambda    
\end{eqnarray}
and noting the meaning of $\Delta_P$, defined by Eq.(\ref{eq: P_j}), concludes that $ -1 \leq \Delta_P \leq 1 $ with
\begin{eqnarray} \label{eq: lam_constraint}
-1 \leq \lambda \leq 1   .
\end{eqnarray}

After these general considerations, we now come back to Eq. (\ref{eq: K}). By setting
\begin{numcases}~
p = 2\sqrt {\frac{ m \hb }{ \uptau(\tau) }} ~ u   \label{eq: u}  \\
q = 2\sqrt {\frac{ m \hb }{ \uptau(\tau) }} ~ v \label{eq: v}  \\
\phi( p ) = e^{iu^2} \varphi(u) \label{eq: phiu} \\
\phi( q ) = e^{iv^2}\varphi(v) \label{eq: phiv}
\end{numcases}~
and using Eq. (\ref{eq: K}), the eigenvalue equation (\ref{eq: eigenvalue}) can be written as
\begin{eqnarray} \label{eq: eigenvalue1}
\frac{1}{\pi} \int_0^{\infty} dv ~ \frac{ \sin(u^2-v^2) }{ u-v } \varphi(v)
&=&  -\lambda \varphi(u)    .
\end{eqnarray}
From this equation, it is clearly seen that the eigenvalues are also independent of mass $m$, Planck constant $\hb$, friction coefficient 
$\ga$ and backflow interval $\tau$. The same is  true for the conservative case.

\subsection{Constant force}

For the linear potential (\ref{eq: lin_pot}), the propagator of CK is given by \cite{MoMi-JPC-2018},
\begin{eqnarray} \label{eq: lin_propag}
G(x, t; x', 0) &=& 
\exp\left[ \frac{i m}{ \hbar} g \left( x ~ \frac{e^{2\ga t}-2\ga t -1}{2\ga(1-e^{-2\ga t})}
+ x' ~ \frac{e^{-2\ga t}+2\ga t -1}{2\ga(1-e^{-2\ga t})} \right)
-\frac{i m}{2\hbar} g^2 ~ \frac{e^{2\ga t}+e^{-2\ga t}-4\ga^2 t^2-2}{2\ga^3(1-e^{-2\ga t})}
\right]
G_{\text{f}}(x, t; x', 0) \nonumber \\
\end{eqnarray}
where $G_{\text{f}}(x, t; x', 0)$ is the propagator (\ref{eq: propagator}) for the free propagation in the CK framework. Following the same steps 
as in the previous subsection, the eigenvalue equation (\ref{eq: eigeneq}) still holds  but now one has that 
\begin{eqnarray} \label{eq: lin_K}
K(p, q) &=&  
-\frac{1}{\pi} \frac{1}{p-q} \exp \left[ i \left( \uptau(\tau) \frac{p^2-q^2}{4m\hb} 
- \frac{(p-q)g}{2\hb} \frac{\uptau(\tau)-\tau}{2\ga} \right) \right]
\sin \left( \uptau(\tau)\frac{ p^2 - q^2 }{4m\hb} - \frac{(p-q)g}{2\hb} \frac{\uptau(\tau)-\tau}{2\ga} \right) 
\end{eqnarray}
with the same property (\ref{eq: K_property}). Then, the eigenvalue equation can be written as 
\begin{eqnarray} \label{eq: lin_eigen}
\frac{1}{\pi} \int_0^{\infty} dv ~ \frac{ \sin[u^2-v^2 - \xi(u-v)] }{ u-v } \varphi(v)
&=&  -\lambda \varphi(u)    .
\end{eqnarray}
where $u$ and $v$ are defined through Eqs. (\ref{eq: u}) and (\ref{eq: v}) but here
\begin{numcases}{}
\phi( p ) = e^{-i( u^2 - \xi u)} \varphi(u) \label{eq: lin_phiu} \\
\phi( q ) = e^{-i( v^2 - \xi v)}\varphi(v) \label{eq: lin_phiv}
\end{numcases}
with
\begin{eqnarray} \label{eq: eta}
\xi &=&  \frac{g}{2} \sqrt{\frac{m}{\hb \uptau(\tau)}} ~ \frac{\uptau(\tau)-\tau}{2\ga}   .
\end{eqnarray}
Note that in the limit $ \ga \rightarrow 0 $ one has $ \xi = - \frac{g}{2} \sqrt{\frac{m}{\hb}}  \tau^{3/2} $ which is the result reported in \cite{BrMe-AP-1998}. Eq. (\ref{eq: lin_eigen}) shows that in the presence of the constant force $m g$, the backflow depends on the parameter $ \xi $ which itself depends on mass, Planck constant, force constant, friction coefficient and the duration of backflow according to Eq. (\ref{eq: eta}).

\subsection{The classical limit}

A quite natural way to study the classical limit is through the so-called quantum-classical transition wave equation originally proposed 
by Richardson et al. in the context of conservative systems \cite{Richardson-2014}. This transition wave equation is governed by a continuous 
parameter covering the two extreme regimes, classical and quantum. The classical one leads to the well-known classical Schr\"odinger equation.
Recently, an extension to open quantum systems has been carried out within the CK framework. We have shown 
\cite{MoMi-AO-2018} that the transition wave equation is expressed as
\begin{eqnarray} \label{eq: CK-e}
i \tilde \hb \frac{\pa }{\pa t} \tilde \psi(x, t)  &=& \bigg[ - e^{-2\ga t} \frac{{\tilde \hb}^2}{2m} \frac{\pa^2}{\pa x^2}  + e^{2\ga t} V(x) \bigg]
\tilde \psi(x, t)  
\end{eqnarray}
where $\tilde \hbar = \hbar \sqrt{\epsilon}$  (scaled Planck constant)  with  $0 \leq \epsilon \leq 1$; $\epsilon = 1$  is for 
the quantum regime and  $\epsilon = 0$ for the classical regime. This parameter gives us an idea of the degree of quantumness of 
the dynamical regime. This equation has also termed by the authors the scaled linear Schr\"odinger equation and the corresponding wave function 
which depends on this parameter can be seen as a transition wave function. With this procedure, the classical limit is reached in 
a continuous way. 

In this context, the scaled probability current density $\tilde j(x, t)$ fulfils the continuity equation
\begin{eqnarray} \label{eq: con_CK-e}
\frac{\pa |\tilde \psi(x, t)|^2}{\pa t} + \frac{\pa \tilde j(x, t)}{\pa x}  &=& 0 ,
\end{eqnarray}
with
\begin{eqnarray} \label{eq: pcd_CK-e}
\tilde j(x, t) &=& \frac{\tilde \hb}{m} \text{Im} \left\{ \tilde \psi^* \frac{\pa \tilde \psi}{\pa x}  \right\} e^{-2\ga t}   .
\end{eqnarray}

The amount of backflow in the time interval $[0, \tau]$  for the scale dynamical regime is then
\begin{eqnarray} \label{eq: P_j-e}
\tilde \Delta_P &\equiv& \tilde P(\tau) - P(0) = - \int_0^{\tau} dt ~ \tilde j(0, t)   .
\end{eqnarray}
Moreover, it is clearly seen that the $K$ operator defined in Eq. (\ref{eq: K}) has also to be rewritten in terms of $\epsilon$, 
$\tilde K$. Then, the corresponding eigenvalue equation (\ref{eq: eigeneq}) also depends on  the scale parameter. However, 
Eq. (\ref{eq: eigenvalue1}) is still independent on the scaled Planck constant. Thus, in the classical limit, interestingly enough, the classical 
Schr\"odinger equation still displays the backflow effect.

\section{The Caldeira-Leggett approach}

In this Section, we are going to introduce the temperature of the environment through the so-called CL equation.
Open quantum systems are usually treated in the framework of the {\it system-plus-environment model} where the total system (the quantum 
system of interest and its environment) is considered to be isolated. In such a model, the total system is represented by a single Hamiltonian and by 
tracing out the degrees of freedom of the environment, the time evolution of the reduced density matrix $\rho$ of the system is
obtained and computed.
In their seminal paper, Caldeira and Leggett \cite{CaLe-PA-1983} modelled the environment as a bosonic bath, an infinite number of quantum 
oscillators in thermal equilibrium, which interacts with the physical system of interest via a position-position coupling. Their model of environment 
is actually a minimal one in the sense that, under certain conditions, reproduces the Brownian motion in the classical regime \cite{Caldeira-book-2014}. 
This approach has been recently used to describe interference and diffraction of identical particles \cite{MoMi-EPJP-2020}.
By tracing out the degrees of freedom of the environment, they derived the following Markovian master equation for the reduced density matrix 
$ \rho $ in the coordinate representation at high-temperatures 
\cite{CaLe-PA-1983, Caldeira-book-2014}
\begin{eqnarray} \label{eq: CL eq}
\frac{\pa \rho}{\pa t} &=& \left[ - \frac{\hb}{2mi} \left( \frac{\pa^2}{\pa x^2} - \frac{\pa^2}{\pa x'^2} \right) - \ga (x-x') \left( \frac{\pa}{\pa x} - \frac{\pa}{\pa x'} \right)
+ \frac{ V(x) - V(x') }{ i\hb }
- \frac{D}{\hb^2} (x-x')^2 \right] \rho(x, x', t)
\end{eqnarray}
where $ D = 2 m \ga k_B T $ plays the role of the diffusion coefficient. One way of solving this equation is by defining the new variables 
$ R = \frac{x+x'}{2} $ and $r = x-x'$, taking the partial Fourier transform with respect to the coordinate $R$ and solving the resulting equation 
by the method of characteristics \cite{Ve-PRA-1994&VeKuGh-PA-1995}. Finally, the inverse Fourier transform is applied to obtain the density 
matrix in the configuration space. From the CL equation (\ref{eq: CL eq}), one can easily reaches the continuity equation,
\begin{eqnarray} \label{eq: con_CL}
\frac{\pa \rho(x, x', t)}{\pa t} \bigg|_{x'=x} + \frac{\pa j(x, t)}{\pa x}  &=& 0 ,
\end{eqnarray}
with 
\begin{eqnarray} \label{eq: pcd_CL}
j(x, t) &=& \frac{\hb}{m} \text{Im} \left\{ \frac{\pa \rho(x, x', t)}{\pa x} \bigg|_{x'=x}  \right\} .
\end{eqnarray}
Diagonal elements of the density matrix are interpreted as probability density.

\subsection{Free evolution of a Gaussian wave packet} \label{ss: CL_single}

The free evolution of the pure state (\ref{eq: Gauss0}) under Eq. (\ref{eq: CL eq}) yields
\begin{eqnarray} \label{eq: rho_t}
\rho(x, t) &\equiv& \rho(x, x, t) = \frac{1}{\sqrt{2\pi} w_t}
\exp \left[
- \frac{ \left( x - x_t \right)^2  }{ 2 w_t^2 }
\right] ,
\end{eqnarray}
for the diagonal elements of the density matrix where
\begin{eqnarray} \label{eq: wt} 
w_t &=& \frac{1}{2\si_p} \sqrt{
	\hb^2 + \frac{ 4\si_p^4 }{ m^2 } \left(\frac{1-e^{-2\ga t}}{2\ga} \right)^2
	+ \frac{ 4\ga t + 4 e^{-2\ga t} - 3 - e^{-4\ga t} }{2 m^2 \ga^3}  \si_p^2 ~ D
} ,
\end{eqnarray}
is the width of the probability density and $x_t$ is given by (\ref{eq: xt}). Comparison of Eq. (\ref{eq: sigmat}) and 
Eq. (\ref{eq: wt}) reveals that the probability density in the CL approach has a temperature dependence through $D$. 
For a given friction $\ga$, the last term under the square root increases with time i.e., 
the width of the probability density for this approach is greater than for the CK one, $w_t > \si_t$. 
Here, one obtains Eq. (\ref{eq: prob_x<0}) for the the 
probability that the particle remains in the half-space $ x<0 $, but with $w_t$ in place of $\si_t$. Since $ \frac{x_t}{\sqrt{2}w_t} $ is not 
{\it essentially} an increasing function of time then here, in the context of CL, the backflow can take place for a single Gaussian wavepacket.
Therefore, the new width $w_t$ of the probability density seems to be responsible for the appearance of backflow with only 
one Gaussian state.

\subsection{Evolution of a superposition of two Gaussian wave packets}

The time evolution of the superposition of two Gaussian wavepackets (\ref{eq: wf_sup_0_Gauss})
\begin{eqnarray} \label{eq: rho0}
\rho_0(x, x') &=& \psi_0(x) \psi_0^*(x')
= N^2 [ \psi_{0a}(x) \psi_{0a}^*(x')  + \al [ e^{-i\theta} \psi_{0a}(x) \psi_{0b}^*(x') + e^{i\theta} \psi_{0b}(x) \psi_{0a}^*(x') ] + \al^2 \psi_{0b}(x) \psi_{0b}^*(x') ]
\nonumber \\
& \equiv &
N^2 [ \rho_{aa}(x, x', 0) + \al ( e^{-i\theta} \rho_{ab}(x,x',0) + e^{i\theta} \rho_{ba}(x,x',0) ) + \al^2 \rho_{bb}(x, x', 0) ]
\end{eqnarray}
under the CL master equation (\ref{eq: CL eq}) can easily be obtained by noting the linearity of this equation. 
One can separately evolve each term 
of the density matrix and then combine the resulting solutions properly i.e., according to Eq. (\ref{eq: rho0}). 
From the procedure outlined at  the beginning of this section, one has that
\begin{eqnarray} \label{eq: rho_ab}
\rho_{ab}(R, r, t) &=& \frac{1}{\sqrt{2\pi} w_t} \exp \left[ a_0(r, t) - \frac{( R - a_1(r, t) )^2}{2w_t^2} \right]
\end{eqnarray}
for the evolution of the cross term $ \psi_{0a}(x) \psi_{0b}^*(x') $ under the CL equation where $w_t$ is given by Eq. (\ref{eq: wt}).
From Eq. (\ref{eq: pcd_CL}), the corresponding probability current density can be now written as
\begin{eqnarray} \label{eq: cur_ab}
j_{ab}(x, t) &=& \frac{\hb}{m} \text{Im} \left\{ 
\left[ \frac{\pa a_0}{\pa r} \bigg|_{r=0} - \frac{x-a_1(0, t)}{2w_t^2} \left( 1 - 2 \frac{\pa a_1}{\pa r} \bigg|_{r=0} \right) \right] \rho_{ab}(x, 0, t)
\right\} 
\end{eqnarray}
%


For free propagation, $V(x) = 0$, we have
\begin{numcases}~
a_0(r, t) = - \frac{(p_{0a}-p_{0b})^2 }{8\si_p^2} 
- \left[  \frac{\si_p^2 e^{-4\ga t}}{2\hb^2} - \frac{1 - e^{-4\ga t} }{4\hb^2\ga} D \right]r^2
+ i ~ e^{-2\ga t} \frac{p_{0a} + p_{0b}}{2 \hb}  r 
\label{eq: a0}
\\
a_1(r, t) = \frac{x_{ta} + x_{tb}}{2}  
+ i \left[ \frac{\hb(p_{0a} - p_{0b})}{4\si_p^2} + \left( \frac{ \si_p^2 e^{-2\ga t} (1-e^{-2\ga t}) }{ 2m\ga \hb } + \frac{ 1 -  e^{-2\ga t}( 2 - e^{-2\ga t}) }{ 4 \hb m \ga^2 } ~ D \right) r \right]     .
\label{eq: a1}
\end{numcases}
In this way, one sees that the probability density, diagonal elements of the density matrix, is obtained from Eq. (\ref{eq: psi2_t}) replacing $ \si_t $ 
by $w_t$. Furthermore, the probability for remaining in the region $ x<0 $ is just given by Eq. (\ref{eq: prob_x<0_CK}) with $w_t$ instead of 
$\si_t$. Note that for $ t< 0 $, the sign of the last term under the square root is negative. Thus, when this term dominates, this leads to an imaginary width which is not acceptable.


Under the presence of the accelerating constant force $ m g $, one simply has
\begin{numcases}~
a_0(r, t) = - \frac{(p_{0a}-p_{0b})^2 }{8\si_p^2} 
- \left[ \frac{\si_p^2 ~ e^{-4\ga t}}{2 \hb^2} - \frac{1 - e^{-4\ga t} }{4\hb^2\ga} D \right]r^2
+ i ~ e^{-2\ga t} \frac{p_{0a} + p_{0b}}{2 \hb}  r 
+ i \frac{m g}{\hb} \frac{1 - e^{-2\ga t} }{2\ga} r
\label{eq: a0_g}
\\
a_1(r, t) = \frac{q_{ta} + q_{tb}}{2}  
+ i \left[ \frac{\hb (p_{0a} - p_{0b})}{4\si_p^2} + \left( \frac{\si_p^2 e^{-2\ga t} (1-e^{-2\ga t}) }{ 2m\ga \hb  } + \frac{ 1 -  e^{-2\ga t}( 2 - e^{-2\ga t}) }{ 4 \hb m \ga^2 } ~ D \right) r \right] 
\label{eq: a1_g}
\end{numcases}
\begin{figure} 
	\centering
	\includegraphics[width=12cm,angle=-0]{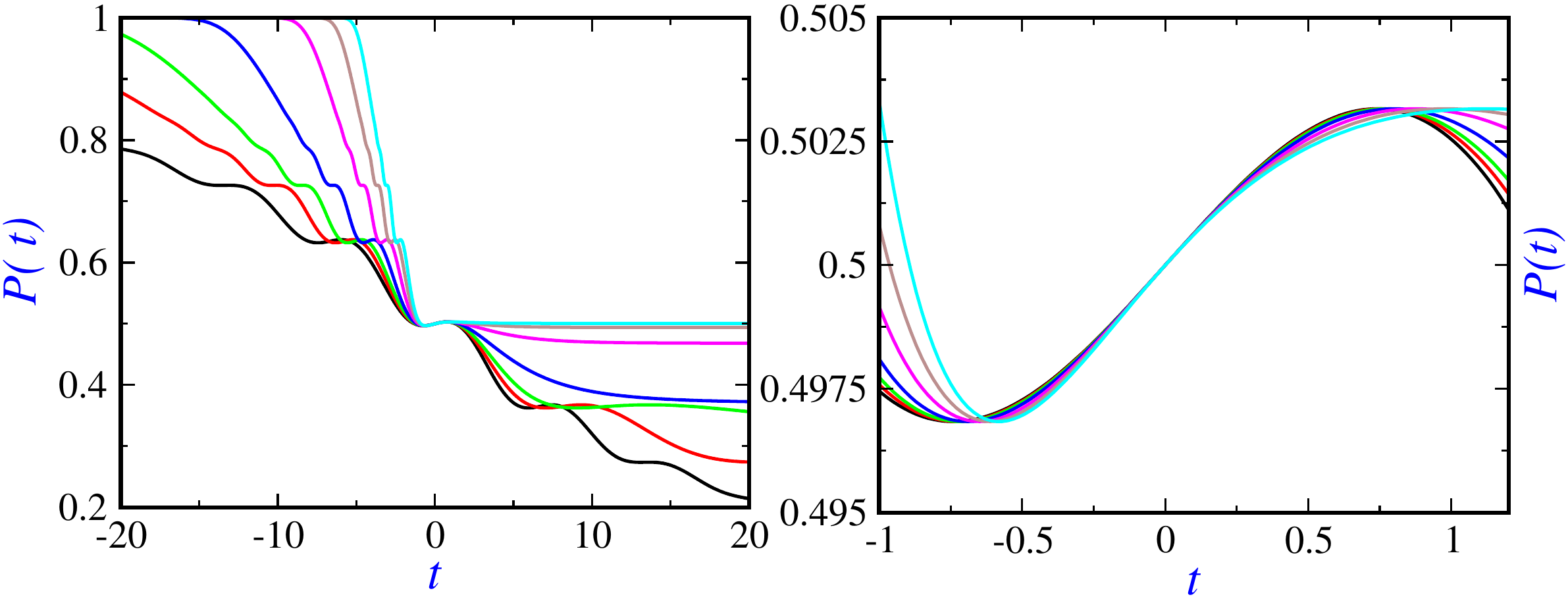}
	\caption{
		The probability for remaining in the region $x < 0$ versus time for the free propagation of the state given by Eq. (\ref{eq: wf_sup_0_Gauss}) in the CK framework (left panel). 
		Different values of the friction coefficient are used: $\ga = 0$ (black), $\ga=0.025$ (red), $\ga=0.05$ (green), $\ga=0.1$ (blue), $\ga=0.2$ (magenta), 
		$\ga=0.3$ (brown) and $\ga=0.4$ (cyan). The right panel is a close-up of the left one around the origin.  
	}
	\label{fig: Probt_CK_free} 
\end{figure}

\section{Numerical calculations}

All numerical calculations in this section are carried out for the initial state given by Eq. (\ref{eq: wf_sup_0_Gauss}) working in a system of units 
where $ \hb = m = 1 $ with $ \si_p = 0.05 $, $p_{0b} = 0.3$, $p_{0a} = 1.4$, $\al = 1.9 $ and $ \theta = \pi $. 
With these values, the amount of backflow is maximal for the non-dissipative dynamics \cite{Albarelli-2016}. 
Thus, from Eq. (\ref{eq: prob_p<0}), the probability for finding the particle with a negative momentum is $ \approx 7.72 \times 10^{-10} $. 
This means that the initial state in the configuration space is practically constructed by superposition of plane waves with positive momenta.

\begin{figure} 
	\centering
	\includegraphics[width=12cm,angle=-0]{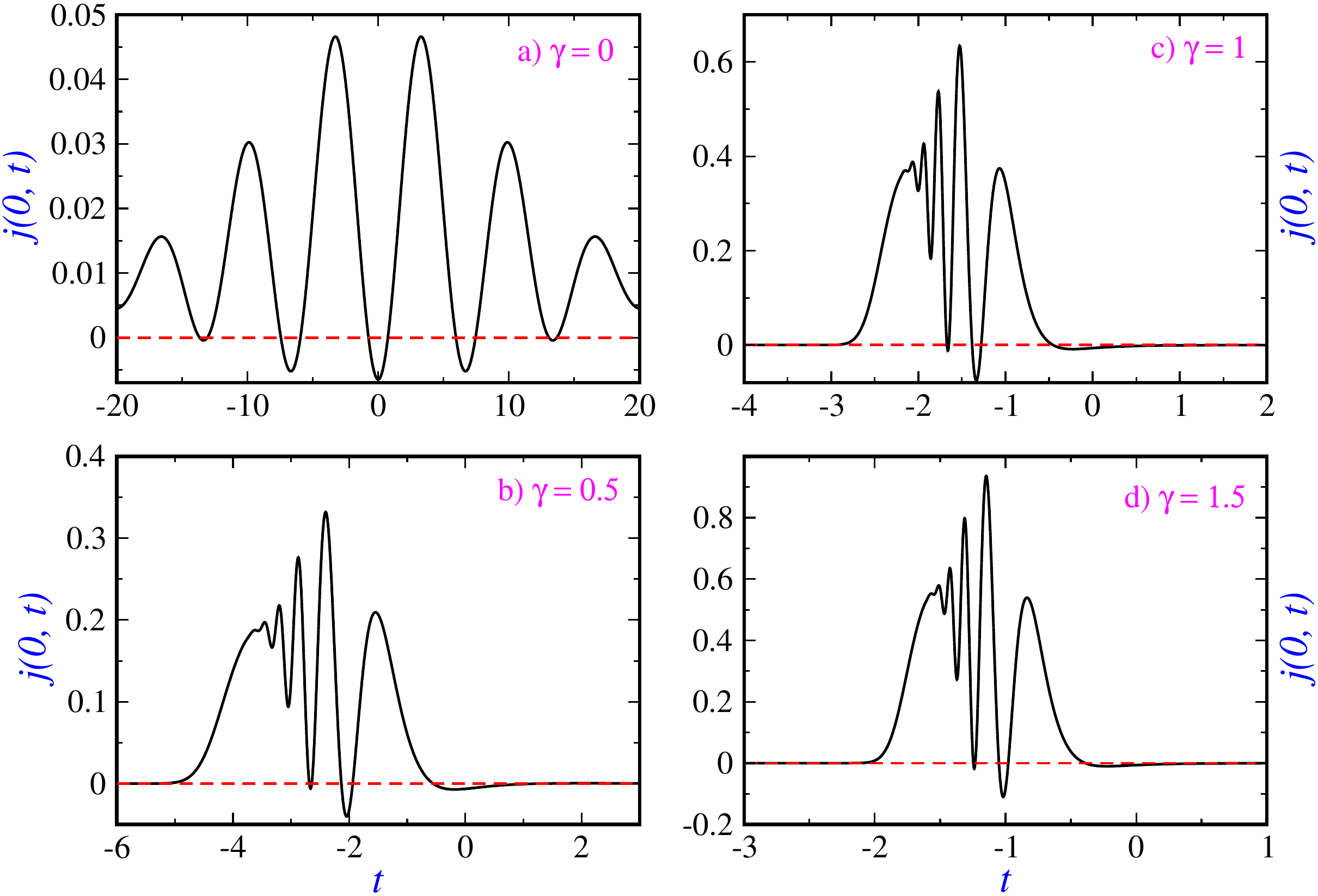}
	\caption{
		The probability current density at the origin versus time, $j(0, t)$, in the CK framework for four different values of friction coefficient. 
	}
	\label{fig: Curt_CK_free} 
\end{figure}

Figure \ref{fig: Probt_CK_free} (left panel) displays the dissipative quantum backflow in the CK framework. 
In this figure, we have plotted the probability for 
remaining in the region $x < 0$ versus time for different values of the friction coefficient: $\ga = 0$ (black), $\ga=0.025$ (red), 
$\ga=0.05$ (green), $\ga=0.1$ (blue), $\ga=0.2$ (magenta), $\ga=0.3$ (brown) and $\ga=0.4$ (cyan). This figure shows that there are some time intervals where the probability for remaining in the 
negative semi-infinite region increases with time. This is the hallmark of the backflow effect. This is better observed in the right panel 
of the same figure where a zoom of the time interval around zero is displayed. As observed, the difference between the  minima and maxima is 
slightly reduced by friction and thus the backflow probability. 

After Eq. (\ref{eq: P_j}), during the backflow intervals where 
the probability increases, the probability current is negative as clearly observed in Figure \ref{fig: Curt_CK_free}
where the probability current density at the origin versus time, $j(0, t)$, for four different values of friction coefficient is plotted. 
\begin{figure} 
	\centering
	\includegraphics[width=12cm,angle=-0]{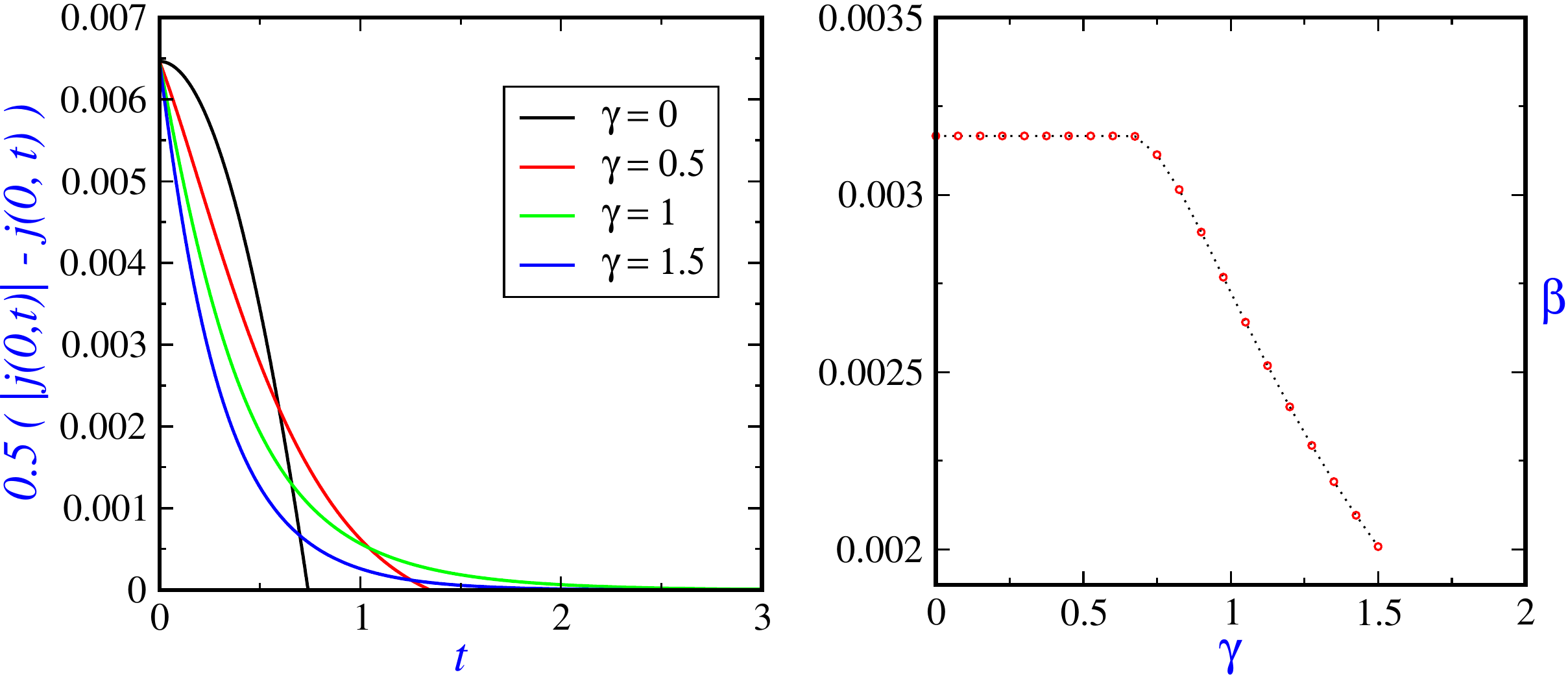}
	\caption{
Negative part of the probability current (left panel) and the backflow amount $\beta$ versus $\ga$ (right panel) in the CK framework. 
	}
	\label{fig: beta_CK_free} 
\end{figure}
According to Eq. (\ref{eq: P_j}), in order to compute the different backflow probabilities, the time intervals which contain 
the negative peaks of the probability current should be identified by finding the zeros of this current. Alternatively, another 
method proposed in \cite{Albarelli-2016} is based on the numerical integration of the negative part of the probability current, 
$ 0.5(|j(0, t)|-j(0, t)) $, over time intervals which contains only a peak.
This quantity is plotted in the left panel of Figure \ref{fig: beta_CK_free} for a period of positive times which contains only the highest 
peak of $ 0.5(|j(0, t)|-j(0, t)) $. 
The maximum amplitude of such a temporary increase of the probability has been used to quantify this effect \cite{Albarelli-2016},  
\begin{eqnarray} \label{eq: betap}
\beta' &=& \underset{t'_1<t'_2}\sup  [P(t'_2) - P(t'_1)]
\end{eqnarray}
where $ (t'_1, t'_2) $ represents the different  time intervals where the backflow occurs. 
In the following we use another criterion to quantify the amount of backflow through the numerical integration of the negative 
part of the probability current over an interval $[t_1, t_2]$ which contains {\it only} its highest peak,
\begin{eqnarray} \label{eq: beta}
\beta &=& \frac{1}{2} \int_{t_1}^{t_2} dt ~ \{ |j(0, t)| - j(0, t) \}.
\end{eqnarray}
Due to the dissipation of energy, to arrive at $x=0$ with velocity $\frac{p_0}{m}$, the condition $  \dot{x}_t \rightarrow + \infty $ must be 
fulfilled at negative times $ t \ll - \ga^{-1} $ for the single Gaussian wavepacket. But, the above condition cannot be fulfilled in practice. Thus, 
we concentrate our attention to positive times. For comparison, the non-dissipative dynamics should also be considered for positive times. 
This means that Eq. (\ref{eq: beta}) should be considered for $ t_1 \geq 0 $.  
Using this method, 
we have computed the backflow amount $ \beta $ for different values of friction coefficient and plotted in the right panel of 
Figure \ref{fig: beta_CK_free}. Interesting enough, this amount decreases with friction after a certain plateau at low friction values. If one allows for negative times then the 
amount of backflow given by Eq. (\ref{eq: beta}) leads to $ \beta \approx 0.006323 $ for the free non-dissipative dynamics.

\begin{figure} 
	\centering
	\includegraphics[width=12cm,angle=-0]{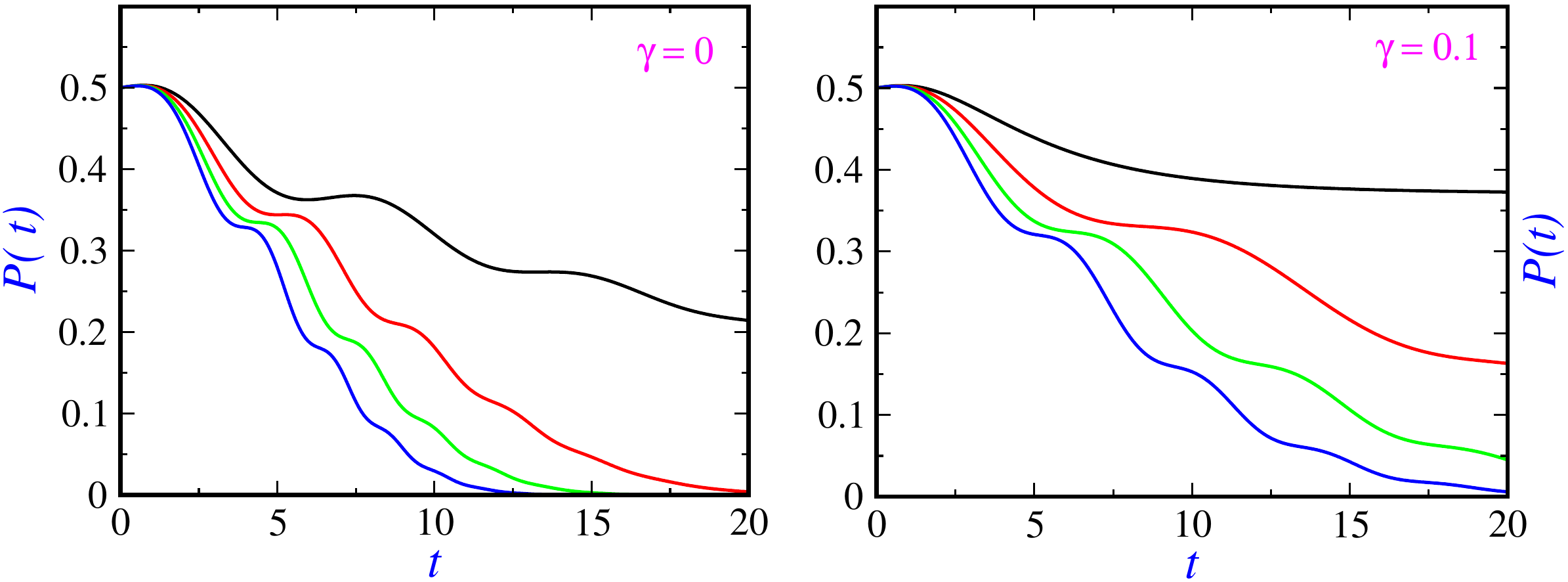}
	\caption{
		The probability for remaining in the region $x < 0$ versus time under the presence of a constant force in the CK framework for the initial 
		state (\ref{eq: wf_sup_0_Gauss}) for two values of the friction coefficient: $\ga = 0$ (left panel) and $\ga=0.1$ (right panel). Different values 
		of the force constant are analyzed: $g=0$ (black curves), $g=0.1$ (red curves), $g=0.2$ (green curves) and $g=0.3$ (blue curves). 
	}
	\label{fig: probt_ck_force} 
\end{figure}

Figure \ref{fig: probt_ck_force} highlights the quantum backflow in the CK framework when a constant force is present. In this figure, 
this effect is studied for non-dissipative and dissipative dynamics separately for different values of the force constant. 
The corresponding backflow probability for remaining in the region 
$x < 0$ versus time under the presence of a constant force for the initial state (\ref{eq: wf_sup_0_Gauss}) and for two values of the 
friction coefficient, $\ga = 0$ (left panel) and $\ga=0.1$ (right panel), is plotted. Different values of the force constant are analyzed:
$g=0$ (black curve), $g=0.1$ (red curve), $g=0.2$ (green curve) and $g=0.3$ (blue curve). As expected, it is seen in both cases, non-dissipative 
and dissipative dynamics, that the backflow probability reduces with the force constant.

\begin{figure} 
\centering
\includegraphics[width=12cm,angle=-0]{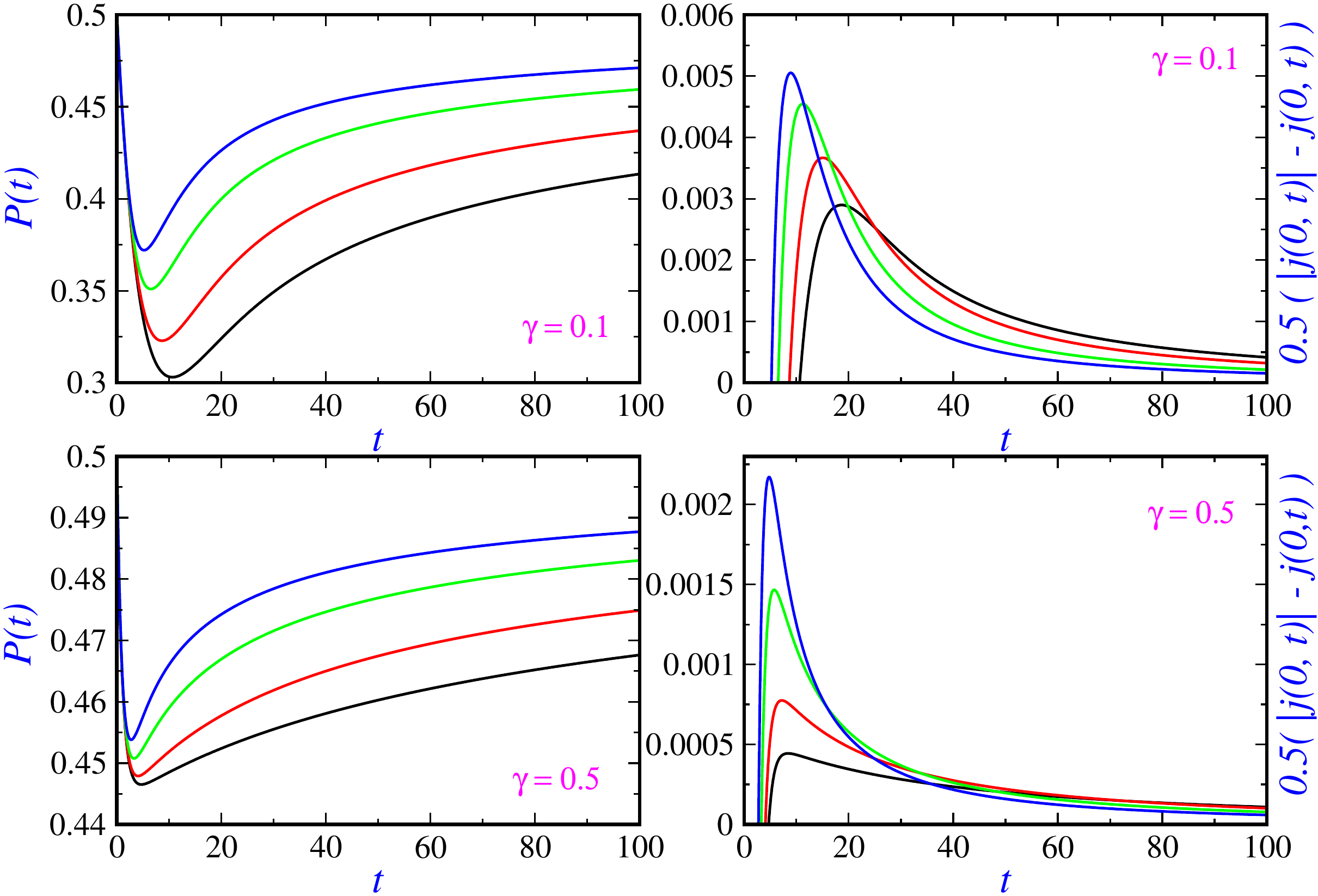}
\caption{
The probability for remaining in $ x < 0 $ versus time in the free propagation of a {\it single Gaussian wave packet}, 
Eq. (\ref{eq: wf_sup_0_Gauss}) with $ \al = 0 $, within the CL framework for two
values of friction coefficients ()$\ga = 0.1$ (top left panel) and $\ga=0.5$ (bottom left panel)) and four values of temperature, 
$k_B T=1$ (black curves), $k_B T=2$ (red curves), $k_B T=5$ (green curves) and $k_B T= 10$ (blue curves). In the right panels, 
we have plotted the negative part of the probability current density $ 0.5( |j(0, t)| - j(0, t) ) $ versus time.
}
\label{fig: singleWP2} 
\end{figure}

In subsection \ref{ss: CL_single}, we have argued that backflow can occur for {\it a single Gaussian wave packet} in the CL approach. 
Occurrence of this backflow has a different nature from the one reported for non-dissipative systems and also dissipative ones in the 
context of CK, where the backflow originates from quantum interference effects. Here, this backflow seems to be originated directly from 
the particular dissipative CL dynamics.
If we consider free evolution of the state (\ref{eq: wf_sup_0_Gauss}) for $\al=0$ in the CL framework then, from Eq. (\ref{eq: rho_t}),
one obtains 
\begin{eqnarray} 
P(t) &=& \int_{-\infty}^0 dx \rho(x, t) = \frac{1}{2} \erfc \left[ \frac{x_{ta}}{\sqrt{2} w_t} \right]
\end{eqnarray}
where $x_{ta}$ is the center of the wave packet given by Eq. (\ref{eq: xt}) replacing $p_0$ by $p_{0a}$, and $w_t$ by Eq. (\ref{eq: wt}).
The width of the wavepacket, $w_t$, increases with time and this increasing width of the probability distribution  works in favour of 
the appearance of backflow. In the CK case, the increasing of the corresponding width, $\si_t$, is countered by the motion of the center 
of the distribution. 
%
In figure \ref{fig: singleWP2} the probability for remaining in $ x < 0 $ (left panels) and the negative part of the probability density 
(right panels) have been plotted versus time for two friction coefficients $\ga = 0.1$ (top panels) and $\ga=0.5$ (bottom panels) and four
temperatures, $k_B T=1$ (black curves), $k_B T=2$ (red curves), $k_B T=5$ (green curves) and $k_B T= 10$ (blue curves). 
This figure clearly shows there is an important backflow interval starting at $t>0$. The corresponding probability increases with time 
reaching apparently a constant value.

\begin{figure} 
\centering
\includegraphics[width=12cm,angle=-0]{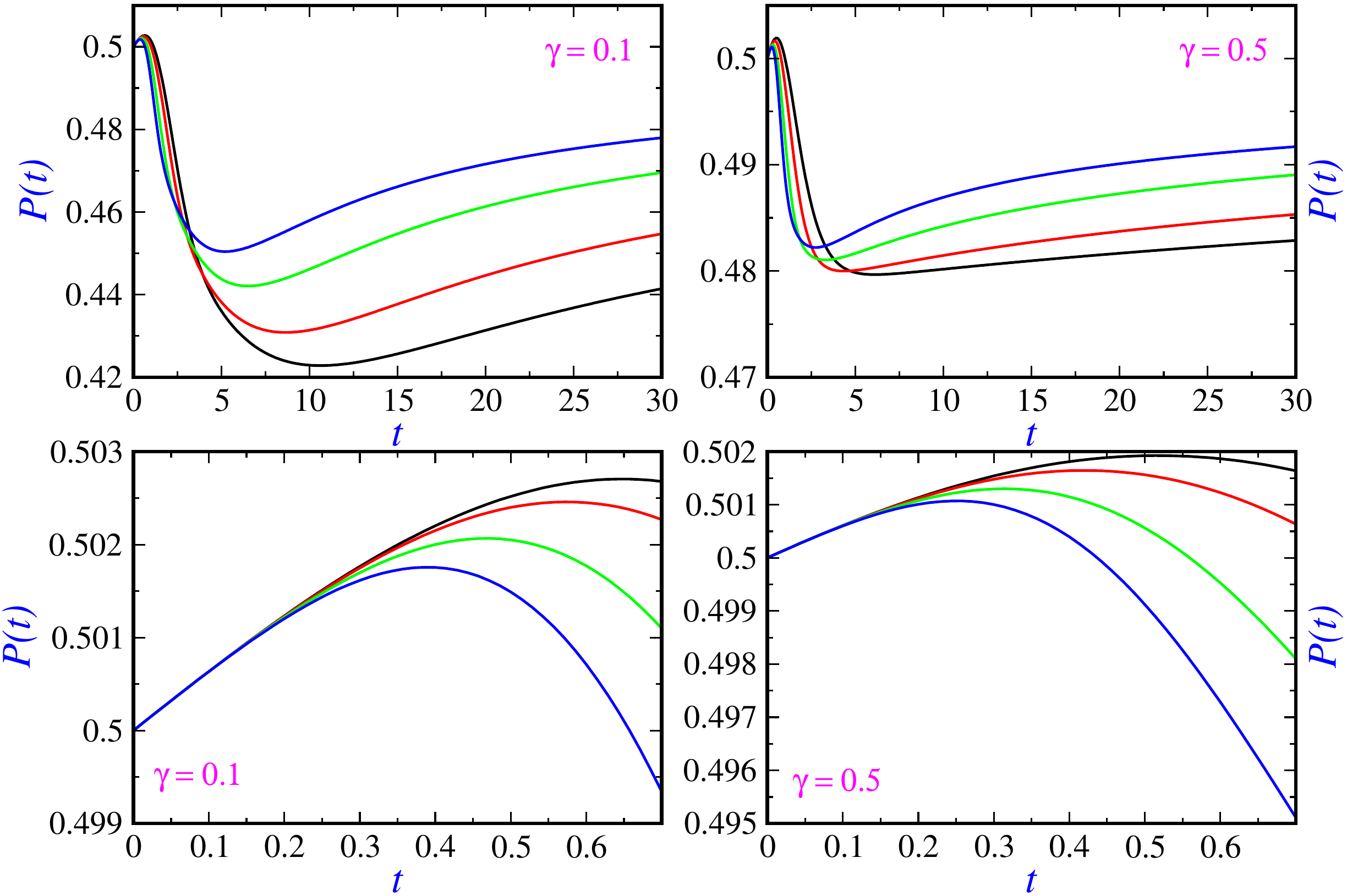}
\caption{
The probability for remaining in $x < 0$ versus time in free propagation of the state (\ref{eq: wf_sup_0_Gauss}) under the CL equation for different 
values of friction coefficients: $\ga = 0.1$ (top left panel) and $\ga=0.5$ (top right panel) for different values of temperature, $k_B T=1$ (black curves),
$k_B T=2$ (red curves), $k_B T=5$ (green curves) and $k_B T= 10$ (blue curves). In the bottom panels we have focused on the first backflow interval.
}
\label{fig: Probt_CL_free} 
\end{figure}

Now we consider the superposition state (\ref{eq: wf_sup_0_Gauss}) in the CL framework.  The influence of the temperature is studied here within the CL framework. In Figure \ref{fig: Probt_CL_free}, the probability of finding the 
particle in the region $x<0$ for free propagation under the CL equation for two values of friction coefficient (0.1 and 0.5) 
and several temperatures ($k_B T=1$ (black curve), $k_B T=2$ (red curve), $k_B T=5$ (green curve) and $k_B T= 10$ (blue curve)) is 
displayed in the top two panels. 
This figure shows two intervals where the mentioned probability increases i.e., the backflow occurs. A close-up of the first time interval  
is seen in the bottom two panels. As it is apparent, the amount and duration of this first backflow decreases with temperature 
for a given friction coefficient. 
Furthermore, for $\ga=0.1$, the duration and amount of the first backflow are respectively 
$\approx 0.657143$ and 
$\approx 0.002705$ for $ k_B T = 1 $ while the corresponding values for $ k_B T = 10 $ are $\approx 0.385714$ and $\approx 0.001756$.  
A similar behavior of the probability is observed at long times with respect to the one Gaussian case.

\begin{figure} 
	\centering
	\includegraphics[width=12cm,angle=-0]{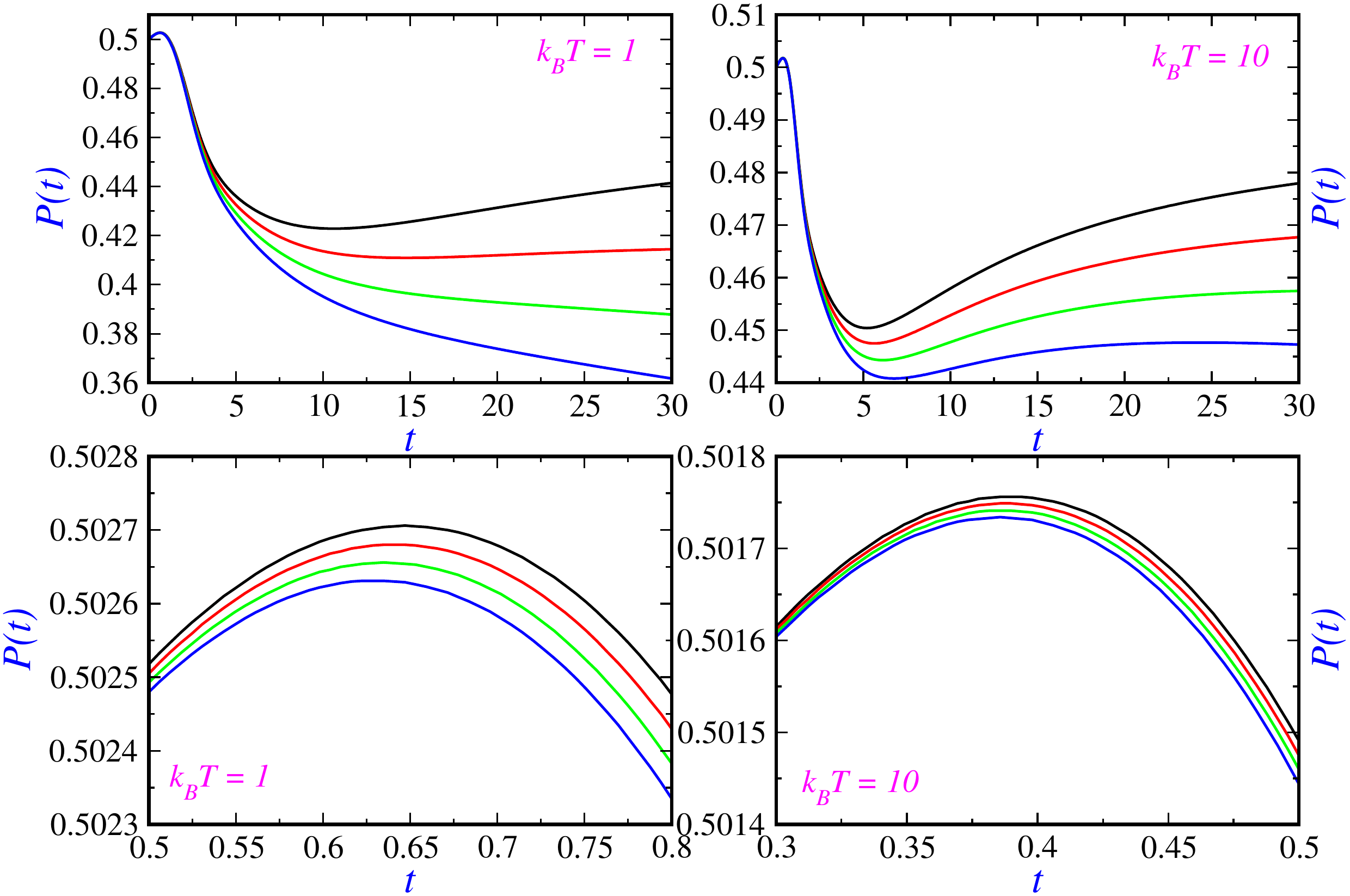}
	\caption{
		The probability for remaining in $x < 0$ versus time in the CL framework for the pure state (\ref{eq: wf_sup_0_Gauss}) for $\ga = 0.1$ 
		for different temperatures: $k_B T = 1$ (top left panel) and $k_B T=10$ (top right panel) for different values of acceleration, $g=0$ 
		(black curves), $g=0.01$ (red curves), $g=0.02$ (green curves) and $g= 0.03$ (blue curves). In the bottom panels we have focused on the 
		region around the maximum.
	}
	\label{fig: Probt_CL_force} 
\end{figure}

Finally, in Figure \ref{fig: Probt_CL_force} is depicted the probability for remaining in the negative part of $x-$axis for dissipative dynamics 
with $\ga = 0.1$ at the temperature $k_B T=1$ (left top panel) and $k_B T=10$ (right top panel) for different values of force constant, 
$g=0$ (black curve), $g=0.01$ (red curve), $g=0.02$ (green curve) and $g= 0.03$ (blue curve). This figure shows again that there is a time interval 
where the probability for remaining in the negative semi-infinite region increases with time. This is better seen in the bottom panels where a 
zoom of this interval is plotted.  
Backflow decreases again with increasing acceleration $g$ for a given temperature. Furthermore, by comparing left and right panels for a 
given $g$, one observes that the temperature decreases the duration and amount of backflow. For instance, when $g=0.03$ 
we have  $ \approx 0.628571 $ and $ \approx 0.002631 $ for $k_B T=1 $ respectively, and $ \approx 0.385714 $ and 
$ \approx 0.001734 $ for $k_B T=10$.

\section{summary and discussion}

In this work, we have studied the backflow effect, that is, negative fluxes for positive-momentum wavepackets for an open quantum system in the CK and CL frameworks. In the CK approach, only friction is taken into account through an effective time-dependent Hamiltonian. In the 
CL approach, the quantum system under study is coupled to a boson reservoir, i.e., to an infinite number of quantum oscillators at thermal 
equilibrium. In this approach, both system and its environment is described by a single Hamiltonian. The equation of motion for the reduced 
density matrix describing the system of interest is obtained by tracing over the degrees of freedom of the environment. In this way,  
Caldeira and Leggett obtained the master equation (\ref{eq: CL eq}) in the high-temperature limit.
In both CK and CL approaches, the equation of motion for the quantum state is linear. 
The other approaches that one could choose are the logarithmic non-linear Schr\"{o}dinger equations \cite{MoMi-AO-2018,
MoMi-EPJP-2019, MoMi-Arxiv-2019} and the stochastic Schr\"odinger equation \cite{Percival,Petruccione}. 
In these approaches, the linearity is not fulfilled and the form of the initial superposed state is not preserved under such dynamics. 

As far as we know, this is the first study of dissipative quantum backflow. For a superposition of Gaussian wavepackets with a 
negligible negative momentum contribution, the backflow also takes place as it has been already reported for the non-dissipative 
dynamics in free propagation. We have found that the backflow decreases with increasing the parameters of the environment i.e., 
friction $\ga$ and temperature $T$ and also observed that it is never suppressed. The constant force $ m g\geq 0 $ behaves 
against backflow. Interestingly enough, quantum backflow is also observed for a single Gaussian wave packet within the CL context.
Surprisingly, the backflow effect seems to be persistent at long times when considering both one and two Gaussian wave packets.
Furthermore, we have studied backflow as an eigenvalue problem in the context of CK framework and concluded 
that eigenvalues are independent of mass, Planck's constant, friction and duration of the backflow in free propagation. However, 
the dynamics under the presence of a constant force is again dependent on those parameters  through a dimensionless quantity. 
Finally, the classical limit has been briefly discussed within a more natural context which is that of the classical Schr\"odinger equation. 
Within this scenario, backflow is not suppressed either.  Quantum backflow for identical particles (fermions and bosons) 
in the conservative and non-coservative cases is now in progress.

\vspace{1cm}
\noindent
{\bf Acknowledgement}
\vspace{1cm}

SVM acknowledges support from the University of Qom and SMA support from 
the Ministerio de Ciencia, Innovaci\'on y Universidades (Spain) under the
Project FIS2017-83473-C2-1-P.  We would like to thank  the referees for providing us very important and critical comments.





\begin{thebibliography} {99}
%
\bibitem{allcock}
G. R. Allcock, Ann. Phys 53 (1969) 253; {\it ibid} 53 (1969) 286; {\it ibid} 53 (1969) 311.	
%
\bibitem{BrMe-JPA-1994}
A. J. Bracken and G. F. Melloy, J. Phys. A Math. Gen. 27 (1994) 2197.
\bibitem{BrMe-AP-1998}
A. J. Bracken and G. F. Melloy, Ann. Phys. (Leipzig) 7 (1998) 726.
%
\bibitem{Penz-2006}
M. Penz, G. Gr\"ubl, S. Kreidl and R. Verch, J. Phys. A 39 (2006) 423.
%
\bibitem{Mu-Lea-PR-2000}
J. G. Muga and C. R. Leavens, Phys. Rep. 338 (2000) 353.
%
\bibitem{Berry-2006}
M. V. Berry and S. Popescu, J. Phys. A:Math. Theor. 39 (2006) 6965.
%
\bibitem{Berry-2010}
M. V. Berry, J. Phys. A: Math. Theor. 43 (2010) 415302.
%
\bibitem{Yearsley-2012}
J. M. Yearsley, J. J. Halliwell, R. Hartshorn and A. Whitby, Phys. Rev. A 86 (2012) 042116.
%
\bibitem{Yearsley-2013}
J. M. Yearsley and J. J. Halliwell, J. Phys.: Conference Serie 442 (2013) 012055.
%
\bibitem{Albarelli-2016}
F. Albarelli, T. Guaita and M. G. A. Paris, Int. J. Quantum Inf., 14 (2016) 1650032.
%
\bibitem{Gu-PRA-2019}
A. Goussev, Phys. Rev. A, 99 (2019) 043626.
%
\bibitem{Ye-PRA-2010}
J. M. Yearsley, Phys. Rev. A 82 (2010) 012116.
%
\bibitem{Caldirola-Kanai}
P. Caldirola Nuovo Cimento 18 (1941) 393; E. Kanai Prog. Theor. Phys. 3 (1948) 440.
%
\bibitem{CaLe-PA-1983}
A. O. Caldeira and A. J. Leggett, Physica A, 121 (1983) 587.
%
\bibitem{Caldeira-book-2014}
A. O. Caldeira, An Introduction to Macroscopic Quantum Phenomena and Quantum Dissipation, Cambridge University Press, 2014.
%
\bibitem{MoMi-AO-2018}
S.V. Mousavi and S. Miret-Art\'es, Ann. Phys. 393 (2018) 76.
%
\bibitem{Sanz-2014}
A. S. Sanz, R. Mart\'{i}nez-Casado, H. C. Pe\~nate-Rodriguez, G. Rojas-Lorenzo and S. Miret-Art\'es, Ann. Phys. 347 (2014) 1.
%
\bibitem{Richardson-2014}
C. D. Richardson, P. Schlagheck, J. Martin, N. Vandewalle and T. Bastin, Phys. Rev. A 89 (2014) 032118.
%
\bibitem{MoMi-JPC-2018}
S.V. Mousavi and S. Miret-Art\'es, J. Phys. Commun. 2 (2018) 035029.
%
\bibitem{MoMi-EPJP-2020}
S.V. Mousavi and S. Miret-Art\'es, Eur. Phys. J. Plus, 135 (2020) 83.
%
\bibitem{Ve-PRA-1994&VeKuGh-PA-1995}
A. Venugopalan, Phys. Rev. A 50 (1994) 2742; A. Venugopalan, D. Kumar and R.
Ghosh, Physica A 220 (1995) 563.
%
\bibitem{Le-LN-2008}
C. R. Leavens: Bohm Trajectory Approach to Timing Electrons, Lect. Notes Phys. 734 (2008)
129–162.
%
\bibitem{MoMi-EPJP-2019}
S.V. Mousavi and S. Miret-Art\'es, Eur. Phys. J. Plus, 134 (2019) 311.
%
\bibitem{MoMi-Arxiv-2019}
S.V. Mousavi and S. Miret-Art\'es, Eur. Phys. J. Plus, 134 (2019) 431.
%
\bibitem{Percival}
I. Percival, {\it Quantum State Diffusion}, Cambridge University Press, Cambridge, 1998.
%
\bibitem{Petruccione}
H.-P. Bauer and F. Petruccione, {\it The Theory of Open Quantum Systems},  Oxford University Press,  2002.
%
\end{thebibliography}
\end{document}